\documentclass[twocolumn,showpacs,preprintnumbers,amsmath,amssymb,aps,superscriptaddress]{revtex4}

\usepackage{graphicx}
\usepackage{dcolumn}
\usepackage{bm}
\usepackage{mathrsfs}
\usepackage{amssymb,amsfonts}
\usepackage{natbib}

\newcommand{\bq}{\begin{eqnarray}}
\newcommand{\eq}{\end{eqnarray}}

\usepackage[dvips]{color}

\begin{document}

\title{Metastable Quantum Phase Transitions in a Periodic One-dimensional Bose Gas: Mean-Field and Bogoliubov Analyses}

\author{R. Kanamoto}
\affiliation{Division of Advanced Sciences, Ochadai Academic Production, Ochanomizu University,
Bunkyo-ku, Tokyo 112-8610 Japan}
\author{L. D. Carr}
\affiliation{Department of Physics, Colorado School of Mines, Golden, CO, 80401, USA}
\author{M. Ueda}
\affiliation{Department of Physics, University of Tokyo, Bunkyo-ku, Tokyo 113-0033 Japan}

\date{\today}


\begin{abstract}
We generalize the concept of quantum phase transitions, which is conventionally defined
for a ground state and usually applied in the thermodynamic limit, 
to one for \emph{metastable states} in \emph{finite size systems}.  In particular, we treat 
the one-dimensional Bose gas on a ring in the presence of both interactions and rotation.  
To support our study, we bring to bear mean-field theory, i.e., the nonlinear Schr\"odinger 
equation, and linear perturbation or Bogoliubov-de Gennes theory.  Both methods give a consistent 
result in the weakly interacting regime: there exist \emph{two topologically distinct quantum phases}.  
The first is the typical picture of superfluidity in a Bose-Einstein condensate on a ring: 
average angular momentum is quantized and the superflow is uniform.  
The second is new: one or more dark solitons appear as stationary states, breaking the symmetry, 
the average angular momentum becomes a continuous quantity. The phase of the condensate can therefore 
be continuously wound and unwound.
\end{abstract}
\maketitle


\section{Introduction}\label{introduction}

One of the main interests of a many-body quantum system is to identify its phase diagram
and quantum phase transitions (QPTs).
The concept of QPTs~\cite{qpt} is usually formulated in the ground state in the thermodynamic limit.
Some criteria for classification of a QPT are the order of transition and the universality class.
Ideally, these can be determined from experimental observables such as energy,
susceptibility, and other bulk measures, and their derivatives with respect to the system
parameter that triggers the transition; more recently, certain forms of entanglement have been
proposed as alternate measures.
A suitably defined order parameter characterizes the symmetry properties of a QPT.
A classical phase transition is driven by thermal fluctuations while a QPT is driven by
quantum fluctuations, and can be most unambiguously understood at zero temperature.

In this article we show that the conventional ways of defining QPTs 
in a Bose-Einstein condensates (BECs) can also be applied to phase transitions in 
\emph{metastable states} in \emph{finite systems}. 
The issue of QPTs in excited states has been addressed previously in nuclear physics~\cite{08:CCI, 08:CS}, 
where nuclei are generically finite and signatures of QPTs can be identified in certain excitation spectra. 
We choose a one-dimensional integrable system of bosons as a concrete example.
Studies of one-dimensional systems relevant to our chosen model have a long history,
including exactly solvable quantum systems~\cite{63:LL,63:L,89:LH-2},
decay of persistent current~\cite{67:LA, 67:Litt, 61:BY, 72:PKU, 73:Blo},
and classical solitons~\cite{89:LH}. 

The most important fact concerning BECs in the thermodynamic limit 
is Hohenberg's theorem~\cite{67:Hoh}, which states that in an interacting and infinite homogeneous system 
condensation does not occur in less than three spatial dimensions at finite temperature and in one dimension at absolute zero. This is specifically proven in the Lieb-Liniger model in the thermodynamic limit~\cite{69:YY}. 
However, this is not true in finite-size systems or spatially confined systems~\cite{91:BK}: 
motional ground-state condensation in two and three dimensions is shown to be possible~\cite{02:LS}. 
Furthermore it is also proved~\cite{02:Fis} via the Bogoliubov inequality that off-diagonal long-range order 
is allowed to exist not only in the ground state, but also in the general excited states. 
The exact theoretical treatment on low-dimensional Bose gases in finite-temperature is 
described for example in Ref.~\cite{04:YC}.

In areas of study such as quantized vortices in superfluid systems, 
metastability plays a key role in the study of quantum dynamics and transport properties. 
This is particularly the case for well-insulated systems, such as ultracold quantum gases,
where there is only negligible exchange of energy and particles with the environment.
In particular, in metastable states of matter waves,
such as soliton trains~\cite{02:ENS-bs,02:Rice-bs}, the effects of dissipation can be suppressed
and a metastable condensate is observable.
The main focus of past studies of QPTs has been on ground states and the thermodynamic limit,
but ultracold quantum gases necessitate a reexamination of the role of excited states.
Moreover, experiments in these systems have precise control over the effective dimensionality, so that
one and two dimensions can be studied for a wide range of interactions~\cite{04:KWW,04:TG,60:Gir},
both repulsive and attractive.

In our previous analysis~\cite{08:KCU} we showed that the average angular momentum
of weakly-repulsive bosons in a one-dimensional ring
undergoes a continuous change in its metastable excited states as a function of interaction and rotation.
This phenomenon is intuitively understood in terms of
bifurcation of stationary excited-state energy branches of the plane-wave state propagating
on the ring, and of localized soliton trains within the mean-field theory.  Each excited state
has a denumerably infinite number of bifurcations from the plane wave to a state containing
1,~2,~$\ldots$ dark solitons; each such bifurcation corresponds to a QPT.
Moreover, for attractive interactions there is a set of QPTs even for the ground state~\cite{03:KSU}. 
For completeness, we compare both repulsive and attractive interactions in our present mean-field study,
although our main focus is on the repulsive case.
In our previous analysis~\cite{08:KCU} we did not discuss in which parameter regime 
and in which energy regime the above phenomenon is observed. 
The aim of the present work is to reveals properties of metastable QPTs within the 
mean-field theory and investigate the linear stability of the mean-field solutions 
in order to show their metastability. We specifically obtain the phase boundary for 
general soliton solutions, which are characterized by the number of density notches 
and phase winding number, in the experimentally variable parameters, i.e., the strength 
of repulsive interaction and the frequency of the rotating drive. The energy diagram 
between the different phases is also shown. In order to show that these excited-state 
solutions do not undergo any dynamical instability, and thereby that the solutions are 
metastable, we investigate the linear stability of all the solutions of the mean-field theory. 

Our presentation is organized as follows. In Sec.~\ref{formulation} we introduce
the Hamiltonian of our system, which corresponds to the Lieb-Liniger model in a 
rotating frame, and describe its basic properties. 
In Sec.~\ref{gp} we derive all stationary states within the mean-field theory
in the weakly interacting regime,
and predict the transition between
the persistent-current state and localized soliton trains.
In Sec.~\ref{Bogoliubov}, we discuss the linear stability of these solutions
in order to determine their metastability against small perturbations.
We summarize our results and discuss experimental possibilities to
realize our ideas in Sec.~\ref{conclusion}.


\section{The Model}\label{formulation}


\subsection{Lieb-Liniger Hamiltonian in a Rotating Frame}\label{LLHam}

Let us consider the Hamiltonian for one-dimensional bosons with a contact interaction,
known as the Lieb-Liniger model~\cite{63:LL}, in position representation,
\bq\label{LLH}
\hat{H}_0 = - \sum_{j=1}^N \frac{\partial^2}{\partial \theta_j^2}
+ g_{\rm 1D}\sum_{j<k}\delta(\theta_j-\theta_k),
\eq
where $N$ is the number of bosonic atoms and $g_{\rm 1D}$ is the effective strength of
$s$-wave interatomic interaction in one dimension (1D)~\cite{98:Ols}.
We impose a periodic boundary condition by assuming a ring-shaped waveguide
or toroidal trap~\cite{03:Dem, 05:GMMPK, 06:AGR, 07:M, 07:NIST-ex, 03:ADW, 08:SBR} of radius $R$.
The coordinates (azimuthal angles) of bosons are specified by a set of variables
$\{\theta\}=\{ \theta_1, \theta_2,\ldots \theta_N \}$ where  $\theta_j\in [0,2\pi)$ for all $j\in\{1,\ldots,N\}$.
The length and energy units are $R$  and $\hbar^2/(2mR^2)$, respectively, and $m$ is the mass of a boson.
The coupling strength $g_{\rm 1D}$ is measured in unit of $\hbar^2/(2mR)$ and hence dimensionless. 
The Hamiltonian we address in this paper is the
Lieb-Liniger Hamiltonian in the rotating frame of
reference,
\bq\label{rotH}
\hat{H}(\Omega)=\hat{H}_0 - 2\Omega \hat{L} + \Omega^2N\,,
\eq
where
\bq
\hat{L}\equiv-i\sum_{j=1}^N\frac{\partial}{\partial \theta_j}
\eq
is the angular-momentum operator measured in units of $\hbar$, the trap rotates at angular frequency $\Omega$, and 
the last term in~(\ref{rotH}) is a constant energy associated with rigid-body rotation which is added to the Hamiltonian to make the system translationally invariant; however it does not change the results of this paper. 

As we will show later, states having angular momentum equals to an integral 
multiple of $N$ are always the stationary solutions (uniform superflow) for repulsive interactions of 
the Hamiltonian~(\ref{LLH}). 
In terms of the nonlinear Schr\"{o}dinger equation approach, there exists another solution having 
a density modulation, so called ``soliton'' solutions. 
In the rest frame, the soliton solutions have node(s) where the density becomes zero, and such a density-modulated 
state with zero-density node(s) is called ``black'' or ``dark'' soliton. 
The uniform-density state and black-soliton state have an finite energy gap and they cannot 
cross over each other at finite strength of repulsive interaction.
By the addition of the rotating-drive term $(-2\Omega\hat{L})$, in contrast, a density-modulated state is 
allowed to have nonzero density notch(es), which is called ``gray'' soliton. This fact makes {\it continuous} 
energy change between distinct topological states, namely, uniform superflow and soliton states possible, 
because the phase modulation of the soliton state can range from infinitesimal (almost constant phase) to 
maximum (corresponding to zero density notches). This is the main finding of our work, and we analytically derive this crossover 
in this paper. 
The terms gray and black are derived from density contrast imagining in  BECs and intensity imaging in optical 
fibers~\cite{agrawal1995}. 
We use the term dark \emph{soliton train} to refer to more than one equally spaced 
soliton, with gray or black indicating whether or not the solitons have nodes. 


\subsection{Periodicity and Umklapp Processes}

All the physical quantities of Hamiltonian~(\ref{rotH}) are periodic with respect to
$\Omega$ with period 1 in our units.
In order to show this, we consider a Schr\"{o}dinger equation as follows~\cite{73:Legg}:
\bq\label{Sch-eq-rot}
\hat{H}(\Omega) \Psi(\{\theta\})={\mathcal E}(\Omega)\Psi(\{\theta\}).
\eq
The many-body wave function $\Psi(\{\theta\})$ satisfies the single-valuedness boundary condition
\bq
\Psi(\theta_1,\ldots,\theta_j,\ldots,\theta_N)=\Psi(\theta_1,\ldots,\theta_j+2\pi,\ldots,\theta_N)
\eq
for all atomic positions $\theta_j$. Substituting the transformation
\bq\label{trs_wf}
\Psi_0(\{\theta\})=\exp\left[-i\Omega \sum_j \theta_j \right]\Psi(\{\theta\})
\eq
into the Schr\"{o}dinger equation~(\ref{Sch-eq-rot}) we can eliminate the $\Omega$-dependent terms
from the equation to yield
\bq\label{new_Sch}
\hat{H}_0\Psi_0(\{\theta\})
={\mathcal E}_0(\Omega)\Psi_0(\{\theta\}).
\eq
The boundary condition of the wave function is then modified to be
\bq\label{bc}
\Psi_0(\theta_1,\ldots,\theta_j,\ldots,\theta_N)\nonumber\\
=e^{2\pi i \Omega}\Psi_0(\theta_1,\ldots,\theta_j+2\pi,\ldots,\theta_N).
\eq
Noting the fact that the new Schr\"{o}dinger equation (\ref{new_Sch}) does not depend on the angular
frequency $\Omega$, and that the boundary condition~(\ref{bc}) is
periodic with respect to $\Omega$ with period of 1,
we observe that the eigenvalue ${\mathcal E}_0(\Omega)$ must be periodic with the same period.
Equation~(\ref{trs_wf}) states that
once the eigensolutions of the Hamiltonian in the rest
frame $\hat{H}_0$ are found, the wavefunction $\Psi(\{\theta\})$ of $\hat{H}$
is obtained via the inverse transformation of Eq.~(\ref{trs_wf}), and
the energy is given by
\bq\label{ene_lab-to-rot}
{\mathcal E} = {\mathcal E}_0- 2\Omega \langle \hat{L} \rangle +\Omega^2 N\,,
\eq
where $\langle \hat{L} \rangle = \langle \Psi_0 |\hat{L}| \Psi_0 \rangle$
is the total angular momentum; this is a conserved quantity and thus a good quantum number.
The periodicity with respect to $\Omega$ enables us to
understand our system in terms of the Bloch energy structure in a solid, although there is not
any direct physical relevance to band theory.
In a solid, the energy dispersion is periodic with respect to quasimomentum and
one can thus define a reduced Brillouin zone in which all information of a system
is included. In our periodic ring, the energy is periodic with respect to $\Omega$ and
one can restrict the analysis within a primitive Brillouin zone $[0,1)$. 
Our restriction of $\Omega \in [0,1)$ for the rest of this study is therefore completely general.

The periodicity with respect to $\Omega$ naturally leads us to define the
{\it Umklapp} process of the total momentum $L'=L+JN$ with $J\in\mathbb{Z}$
an arbitrary integer~\cite{63:LL}. The Umklapp excitations are translations
of the center of mass on the ring.
Namely, a state with angular momentum $L$ has
an infinite number of counterparts at angular momenta separated by integer
multiples of the number of particles, each counterpart having identical properties.


\subsection{Conserved Quantities}\label{Hprop}

The reduced single-particle density matrix of an $N$-body
wave function $\Psi(\{\theta\})$ is given by
\bq
\rho_1 (\theta',\theta)=
\int_0^{2\pi} \!\! d\theta_2 \ldots d\theta_N
\Psi^*(\theta',\theta_2,\ldots,\theta_N)\nonumber\\
\times\Psi(\theta,\theta_2,\ldots,\theta_N),
\eq
The spectral decomposition of $\rho_1$ takes the form
\bq
\rho_1 (\theta',\theta) = \sum_j \lambda_j \psi_j^* (\theta')\psi_j(\theta)\,,
\eq
where $\psi_j$ are the eigenvectors in the many-body Hilbert space and $\lambda_j$ are
the associated eigenvalues.  If there is one and only one dominant eigenvalue of $\rho_1$,
there exists off-diagonal long-range order (a Bose-Einstein condensate, or BEC), and the corresponding
eigenfunction,
\bq
\psi_M(\theta)=|\psi_M(\theta)|e^{i\varphi_M(\theta)}\,,
\label{eqn:condensateWavefunction}
\eq
is regarded as an effective one-body quantity called the \emph{condensate wavefunction.}
Although Bose condensation does not occur in the thermodynamic limit in 1D,
here we consider finite systems for which $N$ does not tend to infinity,
and the concept of BEC in 1D is therefore valid.

The superfluid velocity is defined from the phase of the condensate wavefunction
in Eq.~(\ref{eqn:condensateWavefunction}),
\bq
v_s(\theta)\equiv \frac{\hbar}{m}\frac{\partial}{\partial \theta}\varphi_M(\theta).
\eq
An integral over the superfluid velocity along a closed path is called the {\it circulation},
\bq
C\equiv \int_0^{2\pi}\!\! d\theta \,v_s(\theta),
\eq
which is quantized as
\bq
C=\frac{\hbar}{m}[\varphi(2\pi)-\varphi(0)]=\frac{\hbar}{m}J,
\eq
where $m$ is the mass of the constituent particles in the condensate.

Number, energy, and angular momentum are also conserved in the full quantum theory
according to the usual relations.
In addition, in the mean field theory there is a denumerably infinite set of conserved quantities as
described by Zakharov~\cite{zakharovVE1973}. The original many-body Hamiltonian~(\ref{LLH}) in one dimension is
also integrable, e.g. via the Bethe ansatz. This will be addressed in a later work~\cite{09:KCU}.


\section{Mean-Field Semi-Classical Theory}
\label{gp}


\subsection{Stationary Solutions}\label{meanfield_rep}

When the contact atomic interaction is very weak, specifically,
$g_{\rm 1D}N \lesssim \mathcal{O}(1)$,
the bosons form a condensate, whose static and dynamical properties are described by the
nonlinear Schr\"odinger, or Gross-Pitaevskii (GP) equation.
It is very useful to study the stationary solutions of the GP equation to develop
an intuitive knowledge of the condensate properties.

For convenience we introduce a dimensionless parameter
\begin{equation}
\gamma\equiv \frac{g_{\rm 1D}N}{2\pi}
\end{equation}
that represents the ratio of the mean-field interaction energy to the kinetic energy
(note that our $\gamma$ is not the Tonks-Girardeau parameter).
The GP equation under a rotating drive takes the form
\begin{equation}
\left[\left(-i\frac{\partial}{\partial \theta}-\Omega\right)^2
+2\pi \gamma |\psi(\theta)|^2\right]\psi(\theta)=\mu\psi(\theta),
\label{gpe}
\end{equation}
where $\psi(\theta)$ is the condensate wavefunction from
Eq.~(\ref{eqn:condensateWavefunction}).
Equation~(\ref{gpe}) has two kinds of
solutions~\cite{00:CCR-rep}: plane wave,
\begin{equation}
\psi^{\rm (pw)}_J(\theta)=e^{iJ\theta}/\sqrt{2\pi}\,
\end{equation}
and soliton train,
\begin{equation}\label{solwf}
\psi^{\rm (st)}_{J,j}(\theta)=\sqrt{\rho_j(\theta)}\,e^{i\varphi_{J,j}(\theta)},
\end{equation}
where $J\in\mathbb{Z}$ is the phase-winding number, and $j\in|\mathbb{Z}|$
is the number of density notches in the soliton train.
The derivation of the soliton solutions is summarized in Appendix A.
The amplitude of the soliton-train solution for repulsive interactions, $g_{\rm 1D}>0$, is given by
\bq\label{gray}
\sqrt{\rho_j(\theta)}=
{\mathcal N}\sqrt{1+\eta\ {\rm dn}^2 \left(\frac{jK(\theta-\theta_0)}{\pi},k\right)},
\eq
where dn$(u,k)$ is the Jacobi dn function with elliptic modulus $k \in [0,1]$,
and $K(k)$ and $E(k)$ are elliptic integrals of the first and second kind, respectively.
The parameter $\theta_0$ is an arbitrary coordinate in the interval $0 \le \theta_0 < 2\pi$,
indicating that the soliton solution is a spontaneous-broken-symmetry state.

\begin{figure}[b]
\includegraphics[scale=0.4]{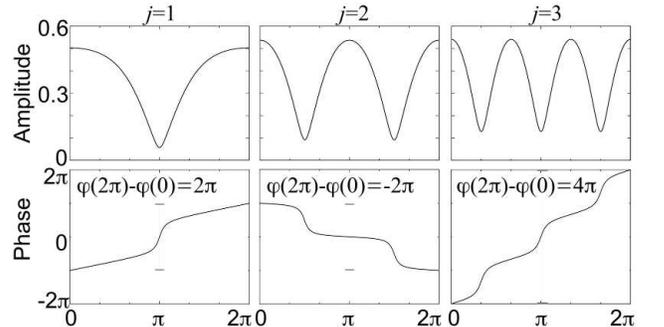}
\caption{
Amplitude $|\psi_{J,j}^{\rm (st)}(\theta)|$ (upper panels) and phase profiles 
$\varphi_{J,j}(\theta)$ (lower panels) of the solitonic condensate wavefunction 
Eq.~(\ref{solwf}) for various numbers of density notches $j$. 
Left: $(j,\gamma,\Omega)=(1,0.7,0.45)$; 
middle: $(j,\gamma,\Omega)=(2,0.7,0.55)$; 
right: $(j,\gamma,\Omega)=(3,0.7,0.45)$. 
For the weakly-interacting regime the phase winding number $J$ is determined 
by a fixed set of parameters $(j,\gamma,\Omega)$. 
}
\label{fig1}
\end{figure}

The normalization constant ${\mathcal N}$ is determined so that the
order parameter is normalized to be unity on the ring:
\bq
{\mathcal N} = \sqrt{\frac{K}{2\pi(K+\eta E)}}\,.
\eq
The depth $\eta$ of density notches is obtained by
substitution of the amplitude into the GP equation as
\bq
\eta=-\frac{2(jK)^2}{g} \in[-1,0].
\eq
We define the functions
\bq\label{fgh}
f &\equiv& \pi^2 \gamma -2 (jK)^2+2 j^2 K E,\\
g &\equiv& f+2 (jK)^2,\\
h &\equiv& f+2(jkK)^2,
\eq
for simplicity.
Then the integration of the imaginary part of the GP equation
gives an analytical expression for the phase prefactor:
\bq
\varphi_{J,j}^{\rm (st)}(\theta)=\Omega\theta-\frac{\cal S}{j K}
\sqrt{\frac{gh}{2f}}\
\Pi\left(\xi; \frac{jK(\theta-\theta_0)}{\pi},k\right),
\eq
where $\Pi(\xi,u,k)$ is an elliptic integral of the third kind with
an amplitude parameter
\bq
\xi\equiv-\frac{2(jkK)^2}{f}
\eq
and a sign function
\bq\label{sign}
{\cal S}\equiv{\rm sign}(\Omega-J)=\left\{
\begin{array}{ccc}
+1 &\quad& \Omega \geq J\\
-1 &\quad& \Omega < J.
\end{array}
\right.
\eq

The soliton-train solution~(\ref{gray}) has two limiting behaviors.
First, in the limit $\eta \to 0$, both the amplitude and phase approach
the plane-wave solution with the same phase winding $J$.
Second, in the limit $\eta \to -1$ (equivalent to $f\rightarrow 0$), the wave function
is found to approach the Jacobi sn function, which corresponds to a black 
soliton train with $\pi$ phase jumps and density notches which form nodes~\cite{00:NIST}.
In this limit $\eta \to -1$, the condensate wave function, chemical potential, and energy are given by

\bq
\psi(\theta)&=&\sqrt{\frac{k^2 K}{2\pi (K-E)}}\ {\rm sn}\left(\frac{jK(\theta-\theta_0)}{\pi},k\right) e^{i\Omega \theta}\,,\\
\mu_{J,j}^{\rm (sn)}&=&\left(\frac{jK}{\pi}\right)^2(1+k^2)\,,\\
{\cal E}_{J,j}^{\rm (sn)}&=&\left(\frac{jK}{\pi}\right)^2 \frac{(1+k^2)E-(1+2k^2)K}{3(E-K)}.
\eq
From analysis of the equation $f=0$ that determines the elliptic modulus,
$k^2$ can be expanded near the critical point as
\bq
k^2 \simeq \frac{4}{j^2}\gamma -\frac{10}{j^4}\gamma^2+O[\gamma^3]\,.
\eq
In between these limits ($-1 < \eta < 0$) we say that Eq.~(\ref{gray}) describes gray solitons. 
In Fig.~\ref{fig1} we show typical amplitude and phase profiles of gray soliton trains for $j=1,2$, and $3$.


\subsection{Phase Diagram}\label{meanfield}

\begin{figure}
\includegraphics[scale=0.41]{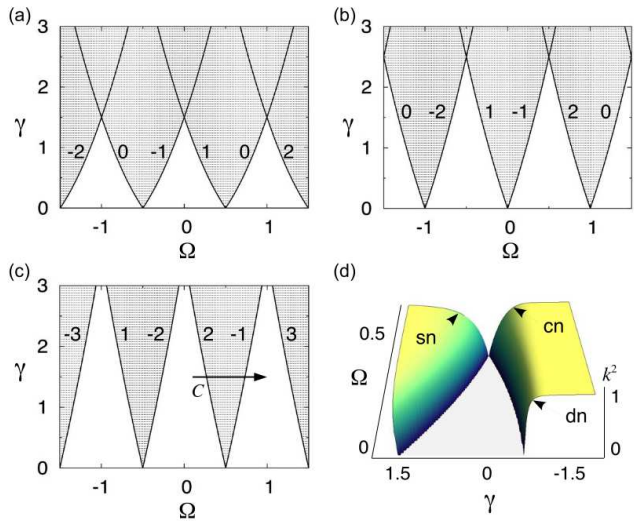}
\caption{(Color online) Parameter space where soliton solutions can exist.
Filled regions present the regimes where the stationary (a) $j=1$, (b) $j=2$, and (c) $j=3$
soliton solutions coexist with plane-wave solutions.
Boundaries are given by the parabolas of Eq.~(\ref{boundary}) with a phase-winding
number $J$, as indicated by an integer value next to the corresponding parabolic curve.
The path $C$ in the panel (c) denotes a typical path to observe the continuous change in the
topology of the order parameter (see, Secs.~\ref{MFene} and \ref{phase}).
(d) Elliptic parameter $k^2$ for a single soliton on the ring ($j=1$) as a function of
rotational drive $\Omega \in [0,0.5]$  and mean field strength $\gamma \in [-1.5, 1.5]$.
For $\Omega=0.5$ the soliton solution is given by the Jacobi sn and cn functions for
repulsive and attractive interactions, respectively. For $\Omega=0$ the soliton solution is
given by the Jacobi dn function for attractive interactions.
}
\label{fig_pb}
\end{figure}

The value of the elliptic modulus $k$ has so far been left undetermined.
To practically obtain the physical quantities such as energy, the density profile and so on,
one needs a concrete value of $k$.
The elliptic modulus is determined from the requirement that the phase of the order parameter
satisfies the single-valuedness condition:
\bq\label{sv}
\varphi_{J,j}(\theta+2\pi)=\varphi_{J,j}(\theta)+2\pi J,
\eq
where the index $J$ physically means the phase-winding number.
Henceforth, we call the condition~(\ref{sv}) the \emph{phase boundary condition for the soliton}.
Equation~(\ref{sv}) can be read as
\bq\label{pbc_r}
2\pi\!\!\!\!\!&&\!\!\!\!\!\!|\Omega-J|\nonumber\\
 &=&2(jk'K)^2\sqrt{\frac{2f}{gh}}
+\sqrt{\frac{2fh}{g}}+j\pi[1-\Lambda_0(\epsilon\backslash \alpha)],
\eq
where
\bq
k'&\equiv&\sqrt{1-k^2}\,,\\
 \epsilon&\equiv&{\rm arcsin}\sqrt{f/h}\,,
 \eq
and $\Lambda_0(\epsilon\backslash\alpha)$ is Heuman's lambda function whose definition
is given in Eq.~(\ref{Heuman}) of Appendix A.
Equation~(\ref{pbc_r}) has a unique real solution
$k \in [0,1]$ if and only if the soliton-train solution exists in the
$(\gamma, \Omega)$-parameter plane. Otherwise only the plane-wave solutions
with an arbitrary phase-winding number exist.
That is, when there is a real solution $k$ of Eq.~(\ref{pbc_r}),
the GP equation~(\ref{gpe}) has the soliton solution in addition
to the plane-wave solution, which is always a formal solution of the GP equation.

For repulsive interactions both the soliton and the plane-wave solutions are stable,
as we demonstrate in Sec.~\ref{Bogoliubov}.
However, for attractive interactions ($\gamma < 0$) the situation is different:
above a certain magnitude of attractive interaction the plane-wave solutions
become dynamically unstable~\cite{00:CCR-att,03:KSU}, and hence the stability of
the plane-wave solution does not hold
for attractive interactions.
The metastable bright soliton-train solutions can be obtained in a similar manner as shown
in Appendix A.

We solve the phase-boundary condition of Eq.~(\ref{pbc_r}) for
$j=1,2,$ and $3$ solitons for repulsive interactions,
and show the solutions in Figs.~\ref{fig_pb}(a) -- \ref{fig_pb}(c), respectively.
In the filled regions the soliton-train solution with $j$ density notches
coexists with plane wave solutions of arbitrary phase winding, and the unfilled region
signifies that the soliton-train solution with $j$ density notches does {\it not} exist.
The phase boundaries written in solid curves will be later identified
as the parabolas given by Eq.~(\ref{boundary}) from the Bogoliubov theory of
Sec.~\ref{Bogoliubov}; each boundary curve corresponds to different value of $J$
in Eq.~(\ref{boundary}). The integer value written next to each parabola in
Fig.~\ref{fig_pb}(a)--(c) represents different values of the phase-winding number $J$.

We can analyze the phase boundaries in Fig.~\ref{fig_pb} as follows.
Two parabolic curves with two distinct phase windings $J$ and $J'$
intersect at certain values of $\Omega$, which we denote by $\Omega_{\rm nodes}(|J-J'|)$,
satisfying $|J'-J|=j$. That is, a soliton state with $j$ density notches and phase winding $J$,
and one with $j$ density notches and phase winding $J'$ become the same sn soliton with $j$
nodes at $\Omega_{\rm nodes}$, provided that the difference in $J$ and $J'$ is equal to $j$.
Along lines $\Omega_{\rm nodes}(|J-J'|)$ for arbitrary $\gamma$,
the phase of the soliton slips by $2\pi |J-J'|$ and the wavefunction is given by the black
soliton train with $j$ density zeros, or nodes.
For instance, for the triple-soliton-train case $j=3$ [Fig.~\ref{fig_pb}(c)],
in the filled region the soliton solution $\psi_{J=2,j=3}^{\rm (st)}$, and
$\psi_{J=-1,j=3}^{\rm (st)}$ exists in $0 < \Omega < 0.5$, and $0.5 < \Omega < 1$, respectively.
There is thus a discontinuous phase jump at $\Omega_{\rm nodes}(|J-J'|) = 0.5$, where
the soliton solution is written as the black soliton train with three density zeros.
The phase-jump lines correspond to
$\Omega_{\rm nodes}(3)\in\{\pm 0.5, \pm 1.5, \ldots\}$ for $j=3$ for arbitrary $\gamma$.
We note that the phase jump also occurs for plane waves at $\Omega_{\rm nodes}$.

Typical behavior of the elliptic modulus $k$, determined by Eq.~(\ref{pbc_r}) for $j=1$,
is shown in Fig.~\ref{fig_pb}(d). The solution $k$ is zero on the phase boundary,
rapidly grows once the parameters $(\gamma, \Omega)$ enter the soliton regime,
and quickly approaches unity. This behavior is qualitatively the same for
$j > 1$ soliton trains.


\subsection{Metastable Quantum Phase Transition}\label{MFene}

We identify the metastable quantum phase transition in the mean field theory, as alluded to
in Sec.~\ref{introduction}.  To do this we study the energy and chemical potential of
all stationary states.  Derivatives of these quantities characterize the order of
the metastable phase transition; our use of the term metastability refers to the fact that we consider
excited as well as ground states.
The energy per particle and chemical potential of the plane-wave state are given by　
\bq\label{uniEC}
{\cal E}_J^{\rm (pw)}&=&(\Omega-J)^2+\frac{\gamma}{2},\\
\mu_J^{\rm (pw)}&=&(\Omega-J)^2+\gamma.
\eq
For soliton solutions the energy per particle and chemical potential are
\begin{widetext}
\begin{eqnarray}
{\cal E}_{J,j}^{\rm (st)}&=&
\gamma+\left(\frac{j}{\pi}\right)^2\left[3KE-(2-k^2)K^2\right]
+\frac{2K^2}{3\gamma}\left(\frac{j}{\pi}\right)^4
\left[3E^2\!-\!2(2-k^2)KE\!+\!K^2(1\!-\!k^2)\right],\label{sol_ene}\\
\mu_{J,j}^{\rm (st)}&=&\frac{3\gamma}{2}+\left(\frac{j}{\pi}\right)^2
\left[3KE-(2-k^2)K^2\right]\,.\label{sol_chem}
\end{eqnarray}
\end{widetext}
The elliptic modulus $k$, which appears both explicitly in Eqs.~(\ref{sol_ene})-(\ref{sol_chem}) and
in the complete elliptic integrals $K=K(k)$ and $E=E(k)$, contains $\Omega$, $\gamma$, $J$, and $j$ implicitly,
as described in Eq.~(\ref{pbc_r}).

\begin{figure}[b]
\includegraphics[scale=0.45]{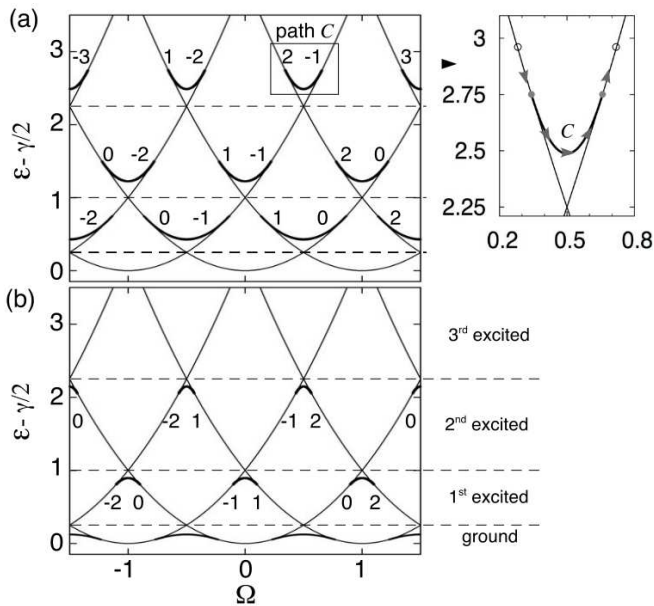}
\caption{
(a) Energy diagram of stationary states for a fixed strength of repulsive ($\gamma=1$) interaction.
The zero of energy is taken as $\gamma/2$. The parabolas correspond to the energies of the plane wave, $E_{J}^{\rm (pw)}-\gamma/2$,
for various phase-winding numbers. Other branches bifurcating from
the parabolas denote the energies of soliton trains, $E_{J,j}^{\rm (st)}-\gamma/2$
for $j=1$ (located in the first-excited-state regime),
$j=2$ (second-excited-state regime),
and $j=3$ (third-excited-state regime) dark solitons, respectively.
Integers in the figures denote the phase-winding number $J$ of solitons, which is
equivalent to that of the plane wave.  
The right panel enlarges the path $C$ which connects the plane-wave and soliton branches in 
$0.3 \lesssim \Omega \lesssim 0.7$ (rough end points are indicated with the open circles), 
starting from the plane-wave branch, passing through 
the soliton branch between 
the phase boundaries (filled circles), and again goes back to the plane-wave branch. 
(b) Energy diagram of stationary states for attractive ($\gamma=-0.4$) interaction.
The $j=1$ bright soliton is located in the ground-state regime, and $j=2$, and $j=3$ bright soliton train
are located in the first, and the second excited regime, respectively.
}
\label{fig_ene}
\end{figure}

The energy diagram shown in Fig.~\ref{fig_ene} summarizes the key result of
mean-field theory.
The figure plots all the stationary-state energies
(choosing the zero of energy to be $\gamma/2$ for plotting convenience) for various
phase-winding numbers $J$ as a function of $\Omega$, with $\gamma$ being fixed.
Taking $\gamma/2$ as the zero of energy removes the trivial $\gamma$ dependence of
${\cal E}_J^{\rm (pw)}$, and we can thus study an excess/shortage of solitonic energy from
the plane-wave energy for a given strength of interaction.
The plane-wave energies are trivial parabolas, and the different values of $J$ (written
as integer values next to the parabolas in the figure)
result in a discrete phase-winding number with respect to $\Omega$.
Let us call the regime ${\cal E}-\gamma/2 <0.5^2$ the ground-state
regime, $0.5^2 \le {\cal E}-\gamma/2 < 1^2$ the first-excited-state regime,
$1^2 \le {\cal E}-\gamma/2 < 1.5^2$ the second-excited-state regime, and so on.

We substitute the solution $k$ of the phase boundary condition of Eq.~(\ref{pbc_r})
for $\gamma=1$ into Eq.~(\ref{sol_ene}), and plot ${\cal E}_{J,j}^{\rm (st)}-\gamma/2$
for various values of $J$ and $j$ in Fig.~\ref{fig_ene}(a).
We find that soliton branches bifurcate from the plane-wave branches
with the same winding number at the phase boundary, so that
\bq
{\cal E}^{\rm (pw)}_J={\cal E}^{\rm (st)}_{J,j},\quad
\partial_{\Omega} {\cal E}^{\rm (pw)}_J = \partial_{\Omega} {\cal E}^{\rm (st)}_{J,j}.
\eq
This shows that a plane wave with phase winding $J$ can be continuously deformed
into a soliton with the same winding number without an energy jump.

Furthermore, at the points $\Omega_{\rm nodes}$,
where the gray soliton train becomes a black soliton train,
the difference in adjacent phase-winding numbers
of solitons equals the number of density notches $j$ of the soliton train
as shown before, and satisfies
\bq
{\cal E}^{\rm (st)}_{J\pm j,j}={\cal E}^{\rm (st)}_{J,j},
\eq
which, again, shows that there is no energy jump associated with
the self-induced phase slip
caused by the soliton.
Similar relations hold for the chemical potential:
\bq
\mu^{\rm (pw)}_J=\mu^{\rm (st)}_{J,j}
\eq
at the phase boundary,
\bq
\mu^{\rm (st)}_{J\pm j,j}=\mu^{\rm (st)}_{J,j},
\eq
at the phase-slip points $\Omega_{\rm nodes}$.  The first derivative
of the chemical potential is a \emph{second-order} cross derivative of the energy,
and is discontinuous:
\bq
\partial_{\Omega} \mu^{\rm (pw)}_J \neq \partial_{\Omega} \mu^{\rm (st)}_{J,j}
\eq
at the phase boundary, where $\mu \equiv \partial{\cal E}/\partial N$.
Therefore the phase transition　is second order.  All of this is true for both　
repulsive and attractive interactions.　


For repulsive interactions [Fig.~\ref{fig_ene}(a)], the ground state is
a plane wave located in the ground-state regime,
with a ground-state phase-winding number
of $J=\lfloor \Omega+1/2 \rfloor$, where $\lfloor x \rfloor$ is the floor function
which gives the integer closest to but not below $x$.
We note that there is no soliton solution in the ground-state regime
and the bifurcations of soliton
branches from the plane-wave branch exist only for excited metastable states.
This is because the total energy increases upon the formation of soliton trains for
repulsive interactions, as the density modulation costs in both the kinetic and interaction energies.
The lowest possible bifurcation thus starts from the first-excited-state regime,
forming the {\it upward}
swallowtail-shaped structure~\cite{swallowtailFootnote} where
these two branches coexist, as can be seen in Fig.~\ref{fig_ene}(a).
The area of this swallowtail vanishes in the noninteracting limit
$\gamma=0$, and the area increases with increasing magnitude of interaction.


Turning to attractive interactions [Fig.~\ref{fig_ene}(b)],
we observe that the density modulation gains in interaction energy.
The soliton branch thus has lower energy than the plane-wave branch,
forming the {\it downward} swallowtail structure, as seen in the figure.
For small attractive interaction of $-0.5 \leq \gamma < 0$ the ground-state branch is
either the plane wave or a single bright soliton state with a ground-state
phase-winding number of $J=\lfloor \Omega+1/2 \rfloor$~\cite{03:KSU-2}--
the two solution types are exclusive.
For soliton solutions in $\gamma < -0.5$, the gain in the interaction energy is always larger than
the loss of the kinetic energy. The soliton branch thus separates away from the plane-wave
branch, and the ground state always becomes a single soliton state ($j=1$).
In a similar manner, the $j=2$ soliton branch, which is located in the
first excited-state regime
in Fig.~\ref{fig_ene}(b), separates away from the parabola for $\gamma < -2$, and the
two lowest-energy states 
are given by $j=1$ and $j=2$ soliton branches.  As shown in~\cite{03:KSU-2},
the bright soliton solution approaches a Jacobi cn function at
$\Omega_{\rm nodes}$, and approaches a Jacobi dn function at the midpoint between
two adjacent node lines $\Omega_{\rm nodes}$.

\begin{figure}[b]
\includegraphics[scale=0.5]{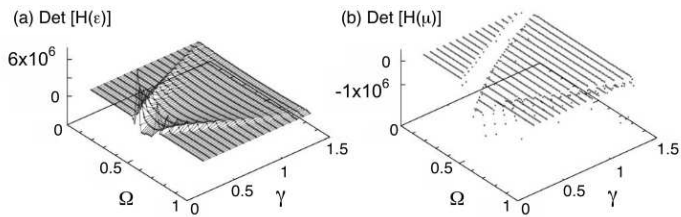}
\caption{
Evidence of a second-order metastable quantum phase transition:
determinant of Hessian matrix for (a) energy and (b) chemical potential.
}\label{fig_detH}
\end{figure}

The continuity of the first and second derivatives of energy with respect to a parameter are
one way to identify the order of a ground-state quantum phase
transition driven by that parameter.
We generalize this idea to the metastable two-parameter QPT
by using the determinant of the Hessian matrix of a function $f(\Omega,\gamma)$,
\bq\label{DetH}
{\rm Det}[H(f)]\equiv \frac{\partial^2 f}{\partial \Omega ^2}\frac{\partial^2 f}{\partial \gamma^2}
-\left(\frac{\partial^2 f}{\partial \Omega \partial \gamma}\right)^2.
\eq
From Eq.~(\ref{uniEC}), ${\rm Det}[H({\cal E}_J^{\rm (pw)})]={\rm Det}[H(\mu_J^{\rm (pw)})]=0$
for arbitrary $(\gamma,\Omega)$ for the plane-wave solutions.
In order to calculate Eq.~(\ref{DetH}) for soliton solutions near the phase boundary,
we numerically calculate the second derivative of energy and chemical potential
for each parameter ($\gamma, \Omega$), and the first cross derivative with respect to both parameters.
Figure~\ref{fig_detH} shows the structure of the determinant of the Hessian for the chemical potential $\mu$,
and energy ${\cal E}$ in the $(\gamma,\Omega)$ plane. The result for the
chemical potential diverges along the phase boundary, while the one for the energy is discontinuous along
the phase boundary. The discontinuity for the latter
increases as $\Omega$ approaches $\Omega_{\rm node}$, and it diverges at $\Omega_{\rm node}$.

\subsection{Phase Winding and Unwinding}\label{phase}
\begin{figure}
\includegraphics[scale=0.43]{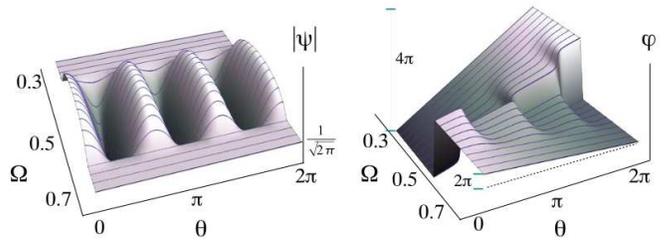}
\caption{(Color online) Change in the amplitude $|\psi(\theta)|$ and phase $\varphi(\theta)$
in the third-excited-state regime for $\gamma=0.6$ for the higher-energy path $C$
shown in Fig.~\ref{fig_ene}.
For $\Omega \simeq 0.3$ the amplitude has a constant value of $|\psi|=1/\sqrt{2\pi}$
and the gradient of the phase is equal to 2.
As $\Omega$ increases, a bifurcation in the energy occurs at $\Omega\simeq 0.39$
and both the amplitude and phase start to wind in the higher-energy
soliton branch while keeping $\varphi(2\pi)-\varphi(0)=4\pi$.
The phase-winding number changes from $J=2$ to $J=-1$ at $\Omega=0.5$ and
the amplitude has three nodes where the phase jumps by $\pi$.
For $\Omega > 0.5$ both the amplitude and phase start to unwind with
$\varphi(2\pi)-\varphi(0)=-2\pi$, and the winding disappears at $\Omega \simeq 0.61$.
The sequence of unwinding through a \emph{single} soliton was described in our previous work~\cite{08:KCU}.
}\label{fig_wf}
\end{figure}

Consider a condensate initially prepared in an excited metastable plane-wave state with a repulsive interaction.
If it takes the continuous higher-energy path of the swallowtail shown in Fig.~\ref{fig_ene}(a) as $\Omega$ is adiabatically changed,
the metastable condensate undergoes an energetically smooth
transition between distinct topological phases through a phase slip at $\Omega_{\rm nodes}$.
Figure~\ref{fig_wf} illustrates the amplitude $\sqrt {\rho(\theta)}$
and phase $\varphi(\theta)$ along the path $C$ $(0.3 \lesssim \Omega \lesssim 0.7)$
indicated in Fig.~\ref{fig_ene}(a) for a fixed repulsive interaction $\gamma=0.6$.
At first the soliton branch does not exist, and the amplitude has a constant value of $1/\sqrt{2\pi}$.
As the rotation $\Omega$ increases, the amplitude starts to modulate at the phase boundary.
The dips in the amplitude deepen as $\Omega$ approaches $\Omega_{\rm nodes}=0.5$.  At $\Omega_{\rm nodes}$
the amplitude develops $j=3$ nodes.  At this point the wave function is described by the Jacobi sn function.
Correspondingly, the gradient of the phase in the plane-wave regime is a constant value $J$,
while the phase starts to wind in the soliton regime, keeping
$\varphi_{2,3}(\theta+2\pi)-\varphi_{2,3}(\theta)=4\pi$.
For the sn soliton at $\Omega=0.5$, the phase jumps by $\pi$ at the node positions.
For $\Omega > 0.5$ the phase boundary condition becomes
$\varphi_{-1,3}(\theta+2\pi)-\varphi_{-1,3}(\theta)=-2\pi$
and the phase is gradually straightened, and gets unwound in the plane-wave regime. 
These changes occur in an energetically continuous manner.


In a superfluid or metastable superflow, the angular momentum is quantized to be
an integral multiple of $N$.
In particular, the ground-state angular momentum $L$ in a weakly repulsive
1D superfluid ring system is quantized as an integral multiple of $N$ at zero temperature,
and there are discontinuous jumps between states having different values of
the phase-winding number.
In fact, this applies only to the ground state, and we show that
the discontinuous jumps are replaced with the continuous crossover
of angular momentum in metastable excited states.

\begin{figure}[b]
\includegraphics[scale=0.35]{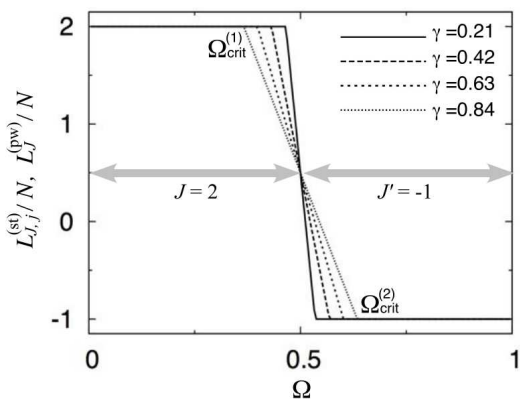}
\caption{
Change in the average angular momentum $L/N$ along path $C$ for a triple soliton train in the
third excited-state regime, where the phase-winding number is $J=2$ $(-1)$ for
$0 \le \Omega < 0.5$ $(0.5 \le \Omega < 1)$.
The average angular momentum takes the value equivalent to the phase-winding number
$J$ in the plane-wave regime, and linearly depends on $\Omega$ in the soliton regime.
}
\label{fig_am}
\end{figure}

The average angular momentum per particle
\bq
\frac{L}{N}\equiv\int_0^{2\pi} d\theta\, \psi^* \left(-i\frac{\partial}{\partial\theta}\right) \psi
\eq
of the plane-wave state is the integer $J$. 
In contrast, $L/N$ of the soliton is noninteger, as can be derived from our mean field formalism:
\bq\label{rep_am}
\frac{L_{J,j}^{\rm (st)}}{N}=
\Omega - \frac{\cal S}{\pi^3 \gamma} \sqrt{\frac{fgh}{2}}\,.
\eq
As the second term becomes zero at the phase boundary,
Eq.~(\ref{rep_am}) coincides with the average angular momentum of plane wave
$L_J^{\rm (pw)}/N=J$.
Figure~\ref{fig_am} plots the average angular momentum $L/N$ along the path $C$ in Fig.~\ref{fig_ene},
for several strength of repulsive interactions. The phase winding and sign ${\cal S}$ are given by
$J=2$ and ${\cal S}=-1$ in $0\le \Omega < 0.5$, and $J'=-1$ and ${\cal S}=1$ in $0.5 \le \Omega < 1$,
respectively.
We denote the \emph{critical angular frequency} in each region by $\Omega^{(1)}_{\mathrm{crit}}$,
and $\Omega^{(2)}_{\mathrm{crit}}$, respectively.

The critical frequencies are determined from linear perturbation theory (Eq.~(\ref{boundary})
in Sec.~\ref{Bogoliubov} below)
by using the values $\gamma$, ${\cal S}$, and the corresponding phase-winding number in each region.
As illustrated in Fig.~\ref{fig_am}, the angular momentum is smoothly connected at $\Omega=\Omega_{\rm nodes}$, and
linearly depends on $\Omega$ in the soliton regime $\Omega^{(1)}_{\mathrm{crit}} < \Omega < \Omega^{(2)}_{\mathrm{crit}}$
with a gradient of
\bq
a=\frac{J'-J}{\Omega^{(2)}_{\mathrm{crit}}-\Omega^{(1)}_{\mathrm{crit}}}\,.
\eq
The angular momentum is thus well fitted by a single line
\bq \frac{L_{J,j}^{\rm (st)}}{N}=a\Omega+b \eq
in $\Omega^{(1)}_{\mathrm{crit}} < \Omega < \Omega^{(2)}_{\mathrm{crit}}$.  We find
\bq\label{rep_am_app}
\frac{L_{J,j}^{\rm (st)}}{N}\simeq\frac{j\Omega-(J+J')\sqrt{\left(\frac{j}{2}\right)^2
+\frac{\gamma}{2}}}{j-2\sqrt{\left(\frac{j}{2}\right)^2+\frac{\gamma}{2}}}\,,
\eq
where $J$ and $J'$ denote the adjacent phase-winding numbers that meet
at $\Omega_{\rm nodes}$, and $b$ is determined by the condition
$a\Omega^{(1)}_{\mathrm{crit}}+b=J$ at the phase boundary.

Thus, we have three critical points in our primitive unit cell for fixed $\gamma (> 0)$:
there is a point at which the higher-energy soliton path begins, $\Omega^{(1)}_{\mathrm{crit}}$.
Part way through this path, a soliton train makes a transition from gray to black, 
at the critical point $\Omega_{\mathrm{nodes}}$, which is always either half-integer or integer and
is independent of $\gamma$.  Finally, the higher-energy soliton path ends at $\Omega^{(2)}_{\mathrm{crit}}$.
A simplified account of this sequence can be found in~\cite{08:KCU}.


\section{Linear Stability of Metastable States}\label{Bogoliubov}

Although all the stationary solutions of the GP equation are listed in
Sec.~\ref{gp}, they may or may not be stable in response to perturbation.
In this section we consider linear perturbation. We show linear stability for the two kinds of
stationary solutions for repulsive interaction by studying fluctuations
around the stationary states of the GP equation, and
argue that the soliton branches in Fig.~\ref{fig_ene} can be realized in practice.


A stationary solution $\psi(\theta)$ of the GP equation under a small perturbation $\delta$
evolves in time as
\bq
\tilde{\psi}(\theta,t)&=&e^{-i\mu t}\{\psi(\theta)\nonumber\\
&&+\sum_n [\delta u_n(\theta) e^{-i\lambda_nt}+\delta v_n^*(\theta) e^{i\lambda_n^* t}]\},
\label{eqn:bdge}
\eq
where $(u_n, v_n)$, and $\lambda_n$ are given as eigenstates and eigenvalues of
the Bogoliubov-de Gennes equations (BdGE), respectively,
and $n\in|\mathbb{Z}|$ denotes the energy-index of excitations.
Recalling the structure of solutions in the Bogoliubov formalism~\cite{01:FS},
for each eigenvalue $\lambda_n$ with positive norm,
\bq\label{BdGnorm}
\int_0^{2\pi}d\theta\,\left[|u_n(\theta)|^2-|v_n(\theta)|^2\right]=1,
\eq
there is also an eigenvalue $\bar{\lambda}_n\equiv -\lambda_{n}$ with negative norm.
The BdGE predict an infinite set of such solutions. One exception to this rule can exist.
This exception corresponds to \emph{Nambu-Goldstone modes}.
For instance, a black soliton is at rest on the ring with respect to any background superflow
in the rotating frame.  The Goldstone mode of the soliton corresponds to 
a center-of-mass translation of the soliton in this frame. 
Henceforth we consider only the Goldstone mode and eigenstates that satisfy~(\ref{BdGnorm}), since the eigenvalues
with negative norm do not have physical meaning.
For the Goldstone mode the corresponding eigenvalue is zero, and the latter eigenstates
that have positive norms can be real (i.e., positive or negative) or complex depending
on the stability of the condensate mode.  In general, $\lambda_n\in\mathbb{C}$.


Let us recall the excitations of a uniform superflow.
When a plane-wave state with phase-winding number $J$
is regarded as a condensate mode, fluctuations from that condensate mode are
given by the eigensolutions of the BdGE with positive norm,
\bq\label{pw_bog}
\lambda_l^{(J,{\rm pw})}=\sqrt{l^2\left( l^2+2\gamma \right)}-2l(\Omega-J), \\
u_l\propto e^{i(J+l)\theta},\quad v_l\propto e^{-i(J-l)\theta},
\eq
where $l\in\mathbb{Z}$
denotes the single-particle angular momentum of the excitation,
which serves as a good quantum number since $[\hat{H}(\Omega),\hat{L}]=0$.

\begin{figure}
\includegraphics[scale=0.45]{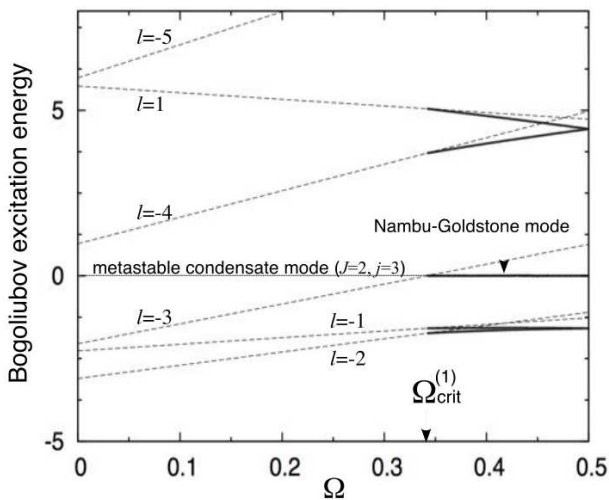}
\caption{Eigenvalues of the BdG equations for a fixed strength of repulsive interaction ($\gamma=1$).
The input condensate mode (zero excitation energy shown in the solid line)
is taken as the higher-energy metastable state in the third excited-state regime of Fig.~\ref{fig_ene}(a).
Dashed and thick curves plot excitation energies from the plane wave and soliton train states, respectively.
}
\label{fig_bog}
\end{figure}

For $\gamma > -0.5$, all the eigenvalues $\lambda_l^{(J,{\rm pw})}$ are real,
as apparent from Eq.~(\ref{pw_bog}).  From Eq.~(\ref{pw_bog}) we also find that
several negative eigenvalues (associated with eigenstates of positive norm)
appear when we take a metastable {\it excited} state as a condensate mode.
These negative eigenvalues correspond to other plane-wave branches located in lower energy regimes
than the input condensate mode itself.
For the case of repulsive interactions, the number of negative eigenvalues thus coincides with the
number of metastable states that are located in a lower energy regime than the metastable state under consideration.
In Fig.~\ref{fig_bog} we show excitation energies with respect to the
plane-wave state with $J=2$ in the third excited-state regime
for $0 \le \Omega \le 0.5$.  For $0.5 \le \Omega \le 1$ the excitation
energies from the plane-wave state with $J=-1$ are symmetric with respect to $\Omega=0.5$.


One of the negative eigenvalues changes its sign at a certain set of parameters
$(\gamma_{\rm crit},\Omega_{\rm crit})$.
 This set of critical values is found by equating $l$ to be ${\cal S}j$ in Eq.~(\ref{pw_bog}) and
imposing the condition$\lambda_{l={\cal S}j}^{(J,{\rm pw})}=0$ as,
\begin{eqnarray}\label{boundary}
\Omega_{\rm crit}-J={\cal S}\sqrt{\left(\frac{j}{2}\right)^2+\frac{\gamma_{\rm crit}}{2}}\,.
\end{eqnarray}
This equation has two real solutions ${\cal S}=\pm 1$.
These solutions determine the phase boundary at which the soliton branch
starts or ends its coexistence with the plane wave.
The region $\gamma \ge \gamma_{\rm crit}$ is identical to requiring that
the phase boundary condition Eq.~(\ref{pbc_r}) has a real solution.


In Fig.~\ref{fig_bog} are shown the Bogoliubov excitation energies from　
the gray soliton branch $\psi_{J=2,j=3}^{\rm (st)}$ along half of the path $C$
indicated in Fig.~\ref{fig_ene}.　
The eigenvalues of the BdGE with the soliton condensate mode
taken as the stationary state $\psi(\theta)$ in Eqs.~(\ref{eqn:bdge})
are also real in the attractive case, so long as $\gamma > -0.5$.
The soliton-train solutions are therefore linearly stable (they are also nonlinearly stable~\cite{carr2000e}
for $\gamma > 0$,
although we do not demonstrate that here, and, according to a quantum tunneling analysis, metastable for $-1<\gamma<0$ with an exponentially long lifetime~\cite{99:UL}, where the difference in the critical point is
associated with the inclusion of the Fock term).
The excitation energies from the soliton branch are found to be close to those from the plane-wave branch.
The notable feature in the soliton regime is that there appears a Nambu-Goldstone mode,
which is continuously connected with one of the negative eigenstates with $l={\cal S}j$ from the plane wave.
This mode reflects the spontaneous symmetry breaking of the soliton-train state.
At the point $\Omega_{\rm nodes}$, a degenerate pair of excitation branches emerges, where the phase jumps up or down by $\pi$ at each soliton in the soliton train.  
For $\Omega > \Omega_{\rm nodes}$ the excitation branches from the soliton train are also symmetric with respect to
$\Omega_{\rm nodes}$.

We proceed to consider the attractive case in more detail.
While all of the excitation energies from the plane-wave state $\lambda_l$ are real for any
$\gamma > -0.5$,
some of the eigenvalues become complex for $\gamma < -0.5$ as seen from Eq.~(\ref{pw_bog}).
Independent of $\Omega$, the excitation energies with $\l=\pm 1$ (i.e., excitation modes of $e^{i(J\pm 1)\theta}$)
become complex for $\gamma <-0.5$, and those with $l=\pm 2$
(i.e., excitation modes of $e^{i(J\pm 2)\theta}$)
become complex for $\gamma <-2$, indicating the modulational instability of
these modes~\cite{mi}.  These complex modes indicate that in
Fig.~\ref{fig_ene}(b)
the $j=1$ soliton branch separates away down from the plane-wave branch in the ground-state regime
for $\gamma <-0.5$, and the soliton branch becomes the ground state while the plane-wave branch
is no longer stable.
Similarly, when the $j=2$ soliton branch separates from the parabolic plane-wave branch, 
it becomes the first excited state; the plane-wave branch has complex $\lambda_l$ and
the plane wave and soliton cannot coexist, in contrast to the repulsive case.

For the potential experimental verification of the continuous crossover between the distinct topological states, 
one may use circular waveguides or toroidal traps to confine a weakly interacting atomic cloud. 
First one has hot atoms above the condensation critical temperature, subjecting to a rotating drive with 
a certain angular frequency. In order to set a system to a metastable uniform superflow, one quickly stops 
the rotation and then lowers the temperature to make the cloud condense. Then the angular frequency of 
the rotating drive should be changed adiabatically. 
In this process, microscopic roughness or small distortion of the trap is enough to create noise sufficient to break 
the translation symmetry of the condensate, making it take the higher-energy path of a metastable soliton 
state, without any artificial distorting the trap. Finally, one stops the adiabatic change in the frequency, 
results in a superflow with a different winding number from the initial state.

Supposing parameters approximately those realized in the experiment of Gupta et al.~\cite{05:GMMPK}, 
using $N=10^5$ $^{87}$Rb atoms in a trap with a transverse frequency $2\pi \times 50$ Hz of the 
circular waveguide with the radius $R=1$ mm, the dimensionless coupling constant with the default three-dimensional scattering length of rubidium is $g_{\rm 1d}=2\pi\times 5.5 \times 10^{-5}$. This results in the mean-field interaction strength $\gamma\simeq 3$. Although this value is about five times larger than the effective strength of interaction where 
the Gross-Pitaevskii type of mean-field approximation is quantitatively valid, the transitions 
between two topologically distinct states yet appear at $\Omega_{\rm cr}$ whose order of magnitude is 
determined by Eq.~(\ref{boundary}). The initial angular frequency of the rotating drive $\Omega$ 
is arbitrary. Once the persistent current with arbitrary $J$, which is determined by the initial 
$\Omega$,  is fixed by stopping the rotation, the number of solitons $j$ is automatically determined by 
Eq.~(\ref{boundary}). 

Summarizing the results of our two theoretical methods, mean field (GPE),
and first-order quantum fluctuations (BdGE) for repulsive interactions,
we have shown that the excitations
from the plane wave in the $j^{\rm th}$ excited-state regime have $j$ thermodynamically unstable modes.
However, there is no modulational instability for repulsive interactions,
and thus both the uniform superflow and gray-soliton states are stable in the $(\gamma,\Omega)$-plane.
At the phase boundary, all energies of the stationary solutions and 
the eigenvalues of the BdG equations continuously connect to the excitations in 
the soliton regime without any energy discontinuity.
In particular, when one of the excitation energies from a plane-wave metastable state
changes the sign from negative to positive, a soliton branch with the same phase-winding number
appears, and the Nambu-Goldstone mode appears as a result.


\section{Conclusions}\label{conclusion}

We have studied metastable excited states of the one-dimensional Bose gas as a function of 
interaction strength and rotation, showing the stability of superflow in a rotating ring trap.
The study of such a system is part of the classic investigation of superfluidity~\cite{leggett1999},
a study to which we have added insight for metastable states.

In the weakly interacting regime, all stationary states and their energy diagrams can be obtained in mean-field theory.  Although it was previously known that the one-dimensional
nonlinear Schr\"{o}dinger equation has both plane-wave and soliton-train solutions on a ring~\cite{00:CCR-rep,00:CCR-att},
we have pointed out that the energy diagram is characterized by the smooth bifurcation of
a soliton branch from a plane-wave branch in the rotating frame.  This is the key to the continuous change in the
topology of the condensate wavefunction~\cite{08:KCU}, which can be characterized as a \emph{self-induced}
phase slip. It is possible to adjust the phase winding and unwinding through the phase slip via a soliton train with nodes,
by rotating the ring trap adiabatically and/or changing the interactions.

For repulsive/attractive interaction, the soliton branch has higher/lower energy 
than the plane-wave branch with the same phase-winding number.  At the phase boundary 
where bifurcations occur, we showed that the excited stationary states undergo a second-order quantum phase transition.

Going beyond mean-field theory, we used Bogoliubov theory to examine the linear stability of
these stationary states, and found that both the plane wave and
soliton train branches are linearly stable for repulsive interactions, and
past a critical value of the interactions, linearly unstable for the attractive case.

The continuous change between topologically distinct states
we have presented in the work can be observed in experiments as follows.
The crossover in the metastable state can be realized starting from hot atoms
confined in a fast-rotating circular waveguide or toroidal trap.
By stopping the rotation and lowering the temperature, one can obtain
a metastable superflow of the condensate of uniform density.
One should then change the angular frequency of the trap adiabatically.
In principle, this trap must have a small deformation
so that the higher-energy path of the swallowtail (see Fig.~\ref{fig_wf}) is selected, but
in practice, no deformation of the trap need be forced on the system, because an infinitesimal perturbation,
as is unavoidable in experiments, is sufficient.
As the angular frequency is increased further across the degeneracy point
where the gray solitons develop nodes and become ``black'', 
the phase-winding number changes over the self-induced phase slip, and
eventually reaches again a superflow of uniform density with a different phase winding from the initial state.
In this way one
is able to observe the phase winding and unwinding without any
abrupt energy discontinuity.

We make the conjecture that the qualitative features of our study and the connection between quantum phase
transitions and semiclassical bifurcations can be found in many quantum field theories.
For instance, we expect that 1D pseudo-spinor bosons on a ring display the same kinds of features,
with the emergence of nonlinear objects in the form of spin-textures signaling a metastable QPT
in the same way that dark solitons did in our scalar theory.  Such features may also appear
in fermionic theories where the energy-gap function takes the place of our mean field.  It is an open question
as to the validity of our concept in higher dimensions, since vortices are fundamentally quantized,
unlike solitons; however, the presence of boundaries may provide for the same kinds of features
as we have described in 1D, since a vortex nucleating on such a boundary can gradually approach
the symmetry axis of a system and thereby increase the average angular momentum continuously.
However, the possibly of a vortex lattice complicates the matter.  Again, in spinor theories
we expect vortex textures to take this role.
Finally, we point out that in the present example both the semiclassical mean-field and the underlying many-body Hamiltonian were integrable.  It would be intriguing to consider an example of metastable QPTs in a system for which one or both of these limits were non-integrable.

This material is based upon work supported by the Sumitomo Foundation (RK), the National Science Foundation under
Grant No. PHY-0547845 as part of the NSF-CAREER program (LDC), and a Grant-in-Aid for Scientific Research (Grant NO. 17071005) (MU).


\appendix
\section{Stationary Solutions of the Nonlinear Schr\"odinger Equation}

In this appendix we provide a detailed derivation of soliton-train solutions of the GP equation
for both repulsive and attractive
interactions.
Substituting the general form of the solution
$\psi(\theta)=\sqrt{\rho(\theta)} e^{i\varphi(\theta)}$ into the GP equation, and
equating the real and imaginary parts respectively, we get
\bq
-\frac{(\sqrt{\rho})''}{\sqrt{\rho}} + \varphi '^2 - 2\Omega \varphi' + \Omega^2
+ 2\pi\gamma \rho =\mu\label{eq1}\,,\\
\varphi'' + 2\varphi'\frac{(\sqrt{\rho})'}{\sqrt{\rho}}
-2\Omega \frac{(\sqrt{\rho})'}{\sqrt{\rho}}=0\,.\label{eq2}
\eq
By integrating Eq.~(\ref{eq1}) we have
\bq
\pi\gamma\rho^3+V \rho
-\left(\frac{\rho'}{2}\right)^2-W^2=\mu \rho^2.
\label{eqn:A3}
\eq
The solution of this equation is given by
\bq\label{amplitude}
\sqrt{\rho(\theta)}=
\left\{
\begin{array}{ll}
{\mathcal N}(\eta) \sqrt{1+\eta\ {\rm dn}^2 \left( \frac{jK(\theta-\theta_0)}{\pi}, k \right) }, & \gamma > 0\\
{\mathcal N}(\eta) \sqrt{{\rm dn}^2 \left( \frac{jK(\theta-\theta_0)}{\pi}, k \right) -\eta k'^2},& \gamma < 0
\end{array}
\right.
\eq
where $k^2+k'^2=1$. The normalization constant ${\mathcal N}$ is determined from
$\int_0^{2\pi} d\theta \rho(\theta)=1$ as
\bq
{\mathcal N}(\eta)=\left\{
\begin{array}{ll}
\sqrt{K/[2\pi(K+\eta E)]^{-1}}, & \gamma > 0\\
\sqrt{K/[2\pi(E-\eta k'^2K)]}, & \gamma <0 .
\end{array}
\right.
\eq
The depth $\eta$ of the density notches is obtained from substitution of
Eq.~(\ref{amplitude}) into Eq.~(\ref{eq1}):
\bq
\eta=\left\{
\begin{array}{lll}
-2(jK)^2/g &\in[-1,0], & \gamma >0\,,\\
g/[2(jk'K)^2] & \in[0,1], & \gamma <0\,.
\end{array}
\right.
\eq
The integral constant $W$ from Eq.~(\ref{eqn:A3}) is given by
\bq
W\equiv\frac{\cal S}{2\pi^4 |\gamma|}\sqrt{\frac{fgh}{2}},
\eq
where we defined functions $f, g, h$, and ${\cal S}$ for notational simplicity as
\bq
f&\equiv&\pm[\pi^2\gamma-2(jK)^2+2j^2KE],\\
g&\equiv& \pi^2\gamma+2j^2KE ,\\
h&\equiv&\pm[\pi^2\gamma-2(jK)^2+2j^2KE+2(jkK)^2],
\eq
where the $\pm$ sign is for repulsive/attractive interactions,
and
\bq
{\cal S}\equiv{\rm sign}(\Omega-J)=
\left\{
\begin{array}{ll}
+1, & \Omega > J,\\
-1, & \Omega < J.
\end{array}
\right.
\eq

From Eq.~(\ref{eq1}) we obtain the chemical potential
\bq
\mu=\frac{3}{2}\gamma+\left(\frac{j}{\pi}\right)^2\left[3KE-(2-k^2)K^2\right].
\label{eqn:cp}
\eq
Note that there is no notational difference between the repulsive and attractive case in Eq.~(\ref{eqn:cp}).
By calculating the interaction energy per particle
\bq
&&{\cal E}_{\rm int}=\pi \gamma\int_0^{2\pi} d\theta \left[\rho(\theta)\right]^2 \nonumber\\
&&=\frac{\gamma}{2}-\frac{2K^2}{3\gamma}\left(\frac{j}{\pi}\right)^4\left[
3E^2 -2 (2-k^2)KE + K^2 (1-k^2)
\right]\nonumber\\
\!\!\!
\eq
one finds the expression for the energy per particle
\bq
{\cal E}_{J,j}^{\rm (st)}&=&
\gamma+\left(\frac{j}{\pi}\right)^2\left[3KE-(2-k^2)K^2\right]\nonumber\\
&+&\!\!\!\frac{2K^2}{3\gamma}\left(\frac{j}{\pi}\right)^4
\left[3E^2\!-\!2(2-k^2)KE\!+\!K^2(1\!-\!k^2)\right], \nonumber\\
\eq
which is applicable to both repulsive and attractive cases.


We next study Eq.~(\ref{eq2}) to obtain the phase prefactor $\varphi(\theta)$ and
rewrite the phase boundary condition
$\varphi(\theta+2\pi)=\varphi(\theta)+2\pi J$.
Equation~(\ref{eq2}) can be readily integrated, giving
\bq
\varphi'(\theta)=\Omega + \frac{W}{\rho}.
\eq
By integrating this equation one more time, the phase part is obtained as
\bq
\varphi^{\rm (st)}=\Omega \theta
+ \frac{\cal S}{j K}\sqrt{\frac{g h}{2f}}\
\Pi\left(\xi; \frac{j K (\theta-\theta_0)}{\pi},k \right), \nonumber\\
\eq
where we used the definition of the elliptic integral of the third kind,
\bq
\Pi(\xi;u,k)=\int du[1-\xi {\rm sn}^2 u]^{-1}.
\eq
Note that the parameter
\bq
\xi \equiv \mp \frac{2(jkK)^2}{f},
\eq
is always positive (negative) for the repulsive (attractive) case.
This difference in the sign is important to rewrite the phase boundary condition.

Since the elliptic integral of the third kind becomes
{\it complete} at $\theta=2\pi$ as
$\Pi(\xi; u, k) = 2j \Pi(\xi \backslash \alpha)$~\cite{abramowitzFootnote},
the phase boundary condition $\varphi(\theta+2\pi)=\varphi(\theta)+2\pi J$
is simplified for the repulsive and attractive cases as
\bq
2 \!\!\!\!\!&&\!\!\!\!\! \pi(\Omega-J) {\cal S}\nonumber\\
&=&\!\!\!\left\{
\begin{array}{ll}
2(jk'K)^2\sqrt{2f/(gh)}+\sqrt{2fh/g}+j\pi\left[1-\Lambda_0(\epsilon \backslash \alpha)\right], \nonumber\\
\sqrt{2gh/f}+j\pi\left[1-\Lambda_0(\epsilon \backslash \alpha)\right],
\end{array}
\right.
\eq
respectively, where
\bq
\epsilon \equiv \left\{
\begin{array}{ll}
\arcsin \sqrt{f/h}, & \gamma > 0 \\
\arcsin \sqrt{h/(k'^2f)}, & \gamma < 0
\end{array}
\right.
\eq
and $\alpha$ is given in terms of the elliptic modulus by $\alpha\equiv{\rm arcsin}(k)$~\cite{abramowitz1964}.
The function $\Lambda_0$ is called \emph{Heuman's lambda function} and defined as
\bq\label{Heuman}
\Lambda_0(\epsilon,k)= \frac{2}{\pi}\left[KE(\epsilon,k'^2)-(K-E)F(\epsilon,k'^2)\right]\!,
\eq
where $F(u,k)$ and $E(u,k)$ are incomplete elliptic integrals of the first and second kinds, respectively.



\begin{thebibliography}{50}
\bibitem{qpt}
S. Sachdev, ``{\it Quantum Phase Transitions}''
(Cambridge University Press, Cambridge, England, 1999).

\bibitem{08:CCI}
M.A.~Caprio, P.~Cejnar, F.~Iachello, 
Ann. Phys. {\bf 323}, 1106 (2008). 

\bibitem{08:CS}
P.~Cejnar, P.~Str\'{a}nsk\'{y}, 
Phys. Rev. E {\bf 78}, 031130 (2008). 

\bibitem{63:LL}
E.H.~Lieb and W.~Liniger, Phys. Rev. {\bf 130}, 1605 (1963).

\bibitem{63:L}
E.H.~Lieb, Phys. Rev. {\bf 130}, 1616 (1963).

\bibitem{89:LH-2}
Y.~Lai and H.A.~Haus, Phys. Rev. A {\bf 40}, 854 (1989).

\bibitem{67:LA}
J.S.~Langer and V.~Ambegaokar, Phys. Rev. {\bf 164}, 498 (1967).

\bibitem{67:Litt}
W.A.~Little, Phys. Rev. {\bf 156}, 396 (1967).

\bibitem{61:BY}
N.~Byers and C.N.~Yang, Phys. Rev. Lett. {\bf 7}, 46 (1961).

\bibitem{72:PKU}
S.J.~Putterman, M.~Kac, G.E.~Uhlenbeck,
Phys. Rev. Lett. {\bf 29}, 546 (1972).

\bibitem{73:Blo}
F.~Bloch, Phys. Rev. A {\bf 7}, 2187 (1973).

\bibitem{89:LH}
Y.~Lai and H.A.~Haus, Phys. Rev. A {\bf 40}, 844 (1989).

\bibitem{67:Hoh}
P.C.~Hohenberg, Phys. Rev. {\bf 158}, 383 (1967). 

\bibitem{69:YY}
C.N.~Yang, and C.P.~Yang, J. Math. Phys. {\bf 10}, 1115 (1969). 

\bibitem{91:BK}
V.~Bagnato, and D.~Kleppner, Phys. Rev. A {\bf 44}, 7439 (1991). 

\bibitem{02:LS}
E.H.~Lieb and R.~Seiringer, Phys. Rev. Lett. {\bf 88}, 170409 (2002).

\bibitem{02:Fis}
U.R.~Fischer, Phys. Rev. Lett. {\bf 89}, 280402 (2002). 

\bibitem{04:YC}
Y.~Castin, J. Phys. IV France, {\bf 116}, 89 (2004). 

\bibitem{02:ENS-bs}
L.~Khaykovich, F.~Schreck, G.~Ferrari, T.~Bourdel, J.~Cubizolles,
L.D.~Carr, Y.~Castin and C.~Salomon, Science {\bf 296}, 1290 (2002).

\bibitem{02:Rice-bs}
K.E.~Strecker, G.B.~Partridge, A.G.~Truscott, and R.G.~Hulet,
Nature {\bf 417}, 150 (2002).

\bibitem{04:KWW}
T.~Kinoshita, T.~Wenger, and D.~Weiss, Science {\bf 305}, 1125 (2004).

\bibitem{04:TG}
B.~Paredes, A.~Widera, V.~Murg, O.~Mandel, S.~F\"{o}lling, I.~Cirac, G.V.~Shlyapnikov,
T.W.~H\"{a}nsch, and I.~Bloch, Nature {\bf 429}, 277 (2004).

\bibitem{60:Gir}
M.~Girardeau, J. Math. Phys. (N.Y.) {\bf 1}, 516 (1960).

\bibitem{08:KCU}
R.~Kanamoto, L.D.~Carr, and M.~Ueda, Phys. Rev. Lett. {\bf 100}, 060401 (2008).

\bibitem{03:KSU}
R.~Kanamoto, H.~Saito, and M.~Ueda,
Phys. Rev. A {\bf 67}, 013608 (2003).

\bibitem{98:Ols}
M.~Olshanii, Phys. Rev. Lett. {\bf 81}, 938 (1998).

\bibitem{03:Dem}
S.O.~Demokritov, A.A.~Serga, V.E.~Demidov, B.~Hillebrands,
M.P.~Kostylev, and B.A.~Kalinikos, Nature {\bf 426}, 159 (2003)

\bibitem{05:GMMPK}
S.~Gupta, K.W.~Murch, K.L.~Moore, T.P.~Purdy, and D.M.~Stamper-Kurn,
Phys. Rev. Lett. {\bf 95}, 143201 (2005).

\bibitem{06:AGR}
A.S.~Arnold, C.S. Garvie, and E. Riis,
Phys. Rev. A {\bf 73}, 041606(R) (2006).

\bibitem{07:M}
S.R.~Muniz, S.D.~Jenkins T.A.B.~Kennedy, D.S.~Naik, and C.~Raman,
Opt. Express {\bf 14}, 8947 (2006).

\bibitem{07:NIST-ex}
C.~Ryu, M.F.~Andersen, P.~Clad\'{e}, V.~Natarajan, K.~Helmerson,
and W.D.~Phillips, Phys. Rev. Lett. {\bf 99}, 260401 (2007).

\bibitem{03:ADW}
B.P.~Anderson, K.~Dholakia, and E.M.~Wright,
Phys. Rev. A {\bf 67}, 033601 (2003).

\bibitem{08:SBR}
G.~Sagu\'e, A.~Badde, and A.~Rauschenbeutal, New J. Phys. {\bf 10}, 113008 (2008).
 
\bibitem{agrawal1995}
G.P.~Agrawal, {\it Nonlinear Fiber Optics} (Academic Press, San Diego, California, 1995).

\bibitem{73:Legg}
A.J.~Leggett, Phys. Fenn. {\bf 8}, 125 (1973).

\bibitem{zakharovVE1973}
V.E.~Zhakharov and A.B.~Shabat, Sov. Phys. JETP {\bf 37}, 823 (1973).

\bibitem{09:KCU}
R.~Kanamoto, L.D.~Carr, and M.~Ueda, to be submitted (2009). 

\bibitem{00:CCR-rep}
L.D.~Carr, C.W.~Clark, and W.P.~Reinhardt,
Phys. Rev. A {\bf 62}, 063610 (2000).

\bibitem{00:NIST}
J.~Denschlagm, J.E.~Simsarian, D.L.~Feder, C.W.~Clark, L.A.~Collins, J.~Cibizolles,
L.~Deng, E.W.~Hagley, K.~Helmerson, W.P.~Reinhardt, S.L.~Rolston, B.I.~Schneider,
and W.D.~Phillips,
Science {\bf 287}, 97 (2000).

\bibitem{00:CCR-att}
L.D.~Carr, C.W.~Clark, and W.P.~Reinhardt,
Phys. Rev. A {\bf 62}, 063611 (2000).

\bibitem{swallowtailFootnote}
The swallowtail described here is a convenient term for a shape in our figure. 
It bears no relation to the swallowtail found nonlinear band theory, 
D. Diakonov, L. M. Jensen, C. J. Pethick, and H. Smith, Phys. Rev. A {\bf 66}, 013604 (2002); 
B.~Wu, R.B.~Diener, and Q.~Niu, Phys. Rev. A {\bf 65}, 025601 (2002); 
B. T. Seaman, L. D. Carr, and M. J. Holland, Phys. Rev. A {\bf 71}, 033622 (2005); 
R. Thom, {\it Structural Stability and Morphogenesis: An Outline of a General Theory of Model} 
(Addison-Wesley, Reading, Massachusetts, 1989). 
However, the general idea of swallowtail forms appearing as a result of bifurcation may imply a deeper connection than is presently understood.

\bibitem{03:KSU-2}
R.~Kanamoto, H.~Saito, and M.~Ueda,
Phys. Rev. A {\bf 68}, 043619 (2003).

\bibitem{01:FS}
A.L.~Fetter and A.A.~Svidzinsky,
J. Phys. Condens. Matter {\bf 13}, R135 (2001).

\bibitem{carr2000e}
L. D. Carr, M. A. Leung, and W. P. Reinhardt, J. Phys. B: At. Mol. Opt. Phys. {\bf 33} 3983 (2000).

\bibitem{99:UL}
M. Ueda and A. J. Leggett, Phys. Rev. Lett. {\bf 83} 1489 (1999).

\bibitem{mi}
A. Hasegawa and W. F. Brinkman, IEEE J. Quantum Electron. {\bf 16} 694 (1980).

\bibitem{leggett1999}
A. J. Leggett, Rev. Mod. Phys. {\bf 71}, S318 (1999).

\bibitem{abramowitzFootnote}
See Ref.~\cite{abramowitz1964} for an explanation of this notation.

\bibitem{abramowitz1964}
M. Abramowitz and I. A. Stegun, \emph{Handbook of Mathematical Functions}
(National Bureau of Standards, Washington, D.C., 1964).

\end{thebibliography}
\end{document}